\documentclass{article}[10pt]

\usepackage{bbm,times,amsmath}

\begin{document}


\bigskip\bigskip\bigskip

\noindent {\LARGE{\bf Gaussian quantum channels}}
\bigskip\bigskip

\noindent {{J.\ Eisert$^{1,2}$ and M.M.\ Wolf$^3$}\vspace{7pt} }

\noindent
{\footnotesize{\it 
\renewcommand{\baselinestretch}{0.75}
$^1$ Blackett Laboratory\\
Imperial College London\\
London SW7 2BW, UK\smallskip\\
$^2$ Institute for Mathematical Sciences\\
Imperial College London\
Exhibition Road\\
London SW7 2BW, UK\smallskip\\
$^3$ Max-Planck-Institut f{\"u}r Quantenoptik\\
Hans-Kopfermann-Stra{\ss}e 1\\
85748 Garching, Germany}}\\



\newcommand{\R}{\mathbbm{R}}
\newcommand{\rr}{\mathbbm{R}}
\newcommand{\nn}{\mathbbm{N}}
\newcommand{\cc}{\mathbbm{C}}
\newcommand{\id}{\mathbbm{1}}

\newcommand{\tr}{{\rm tr}\,}
\newcommand{\gr}[1]{\boldsymbol{#1}}
\newcommand{\be}{\begin{equation}}
\newcommand{\ee}{\end{equation}}
\newcommand{\bea}{\begin{eqnarray}}
\newcommand{\eea}{\end{eqnarray}}
\newcommand{\ket}[1]{|#1\rangle}
\newcommand{\bra}[1]{\langle#1|}
\newcommand{\avr}[1]{\langle#1\rangle}
\newcommand{\G}{{\cal G}}
\newcommand{\eq}[1]{Eq.~(\ref{#1})}
\newcommand{\ineq}[1]{Ineq.~(\ref{#1})}
\newcommand{\sirsection}[1]{\section{\large \sf \textbf{#1}}}
\newcommand{\sirsubsection}[1]{\subsection{\normalsize \sf \textbf{#1}}}

\newcommand{\proofend}{\hfill\fbox\\\medskip }

\medskip

\noindent
\begin{minipage}{.9\textwidth}
{\small{This article provides an elementary 
introduction to Gaussian channels
and their capacities. We review  results on the classical, quantum,
and entanglement assisted capacities and discuss related entropic
quantities as well as additivity issues. Some of the known results
are extended. In particular, it is shown that the quantum
conditional entropy is maximized by Gaussian states and that some
implications for additivity problems can be extended to the
Gaussian setting.}}
\end{minipage}
\medskip

\tableofcontents

\bigskip\bigskip

\section{Introduction}

Any physical operation that reflects the time evolution of the
state of a quantum system can be regarded as a channel. In
particular, quantum channels grasp the way how quantum states are
modified when subjected to noisy quantum communication lines.
Couplings to other external degrees of freedom, often beyond
detailed control, will typically lead to losses and decoherence,
effects that are modelled by appropriate non-unitary quantum
channels.

Gaussian quantum channels play a quite central role indeed.
After all, good models
for the transmission of light through fibers are provided by
Gaussian channels. This is no accident: linear couplings of
bosonic systems to other bosonic systems with quadratic
Hamiltonians can in fact appropriately  be said to be ubiquitous
in physics. In this optical context then, the time evolution of
the modes of interest, disregarding the modes beyond control, is
then reflected by
 a Gaussian bosonic channel. Random classical noise,
introduced by Gaussian random displacements in phase space, gives
also rise to a Gaussian quantum channel, as well as losses that
can be modelled as a beam splitter like interaction with the
vacuum or a thermal mode.

This article provides a brief introduction into the theory of
Gaussian quantum channels. \footnote{This is a review article.
Previously unpublished material is presented in Section 3.3 and in
Section 5.4.} After setting the notation and introducing to the
elementary concepts, we provide a number of practically relevant
examples. Emphasis will later be put on questions concerning
capacities: Capacities come in several flavours, and essentially
quantify the usefulness of a quantum channel for the transmission
of classical or quantum information. We will briefly highlight
several major results that have been achieved in this field.
Finally, we discuss a number of open questions, notably related to
the intriguing but interesting and fundamental questions of
additivities of quantum channel capacities.

\section{Gaussian channels}\label{gauss}

In mathematical terms a {\it quantum channel} is a completely
positive trace-preserving map $\rho \longmapsto T(\rho)$ that
takes states, i.e., density operators $\rho$ acting on some
Hilbert space ${\cal H}$, into states\footnote{This expression
refers to the Schr{\"o}dinger picture of quantum channels.
Equivalently, one can define the dual linear map $T^\ast$ in the
Heisenberg picture via $\text{tr}[\rho T^\ast (A)]=
\text{tr}[T(\rho)A]$, which in turn is then completely positive
and unital.}. For simplicity we will always assume that output and
input Hilbert spaces are identical. Every channel can be conceived
as reduction of a unitary evolution in a larger quantum system. So
for any channel $T$
there exists a state $\rho_E$ on a Hilbert space ${\cal H}_E$, and
a unitary $U$ such that
\begin{equation}\label{Channel}
    T(\rho) = \text{tr}_E[ U (\rho\otimes \rho_E) U^\dagger].
\end{equation}
The system labeled $E$ serves as an environment, embodying degrees
of freedom of which elude the actual observation, inducing a
decoherence process. The channel is then a local manifestation of
the unitary evolution of the joint system. A {\it Gaussian
channel} \cite{Channels,capa,HolWe,OurOldChannels,Lindblad} is now
a channel of the form as in Eq.\ (\ref{Channel}), where $U$ is a
Gaussian unitary, determined by a quadratic bosonic Hamiltonian,
and $\rho_E$ is a Gaussian state \cite{Holevobook}. In many cases,
of which the lossy optical fiber is the most prominent one, this
restriction to quadratic Hamiltonians gives a pretty good
description of the physical system. Note that although the channel
is assumed to be Gausssian in the entire article, the input states
are not necessarily taken to be Gaussian.

\subsection{Preliminaries}

It seems appropriate for the following purposes to briefly fix the
notation concerning Gaussian states and their transformations
\cite{Lindblad,Holevobook,book,Art}. For a quantum system with $n$
modes, i.e., $n$ canonical degrees of freedom, the {\it canonical
coordinates} will be denoted as $R=(x_1, p_1,...,x_n,p_n)$. Most
naturally, these operators can be conceived as corresponding to
field quadratures. Although all statements in this article hold
true for any physical system having canonical coordinates, we will
often refer to the optical context when intuitively describing the
action of a channel. The creation and annihilation operators are
related to these canonical coordinates according to $x_i =( a_i +
a_i^\dagger)/\sqrt{2}$ and $p_i=-i(a_i - a_i^\dagger)/\sqrt{2}$.
The coordinates satisfy the canonical commutation relations, which
can be expressed in terms of the {\it Weyl operators} or {\it
displacement operators}
    $W_\xi = e^{ i \xi^T \sigma R}$
with $\xi \in \rr^{2n}$:
\begin{equation}
    W_\xi^\dagger W_{\xi'} =W_{\xi'} W_{\xi}^\dagger
     e^{i \xi^T \sigma \xi'},\,\,\,\sigma = \bigoplus_{i=1}^n
    \left(
    \begin{array}{cc}
    0 & 1\\
    -1 & 0 \\
    \end{array}
    \right),
\end{equation}
where we have set $\hbar=1$. The matrix $\sigma$ defines the
symplectic scalar product, simply indicating that position and
momentum of the same mode do not commute.

The Fourier transform of the ordinary Wigner function in phase
space $\rr^{2n}$ is the {\it characteristic function} \be\chi_\rho
(\xi) = {\rm tr}[ \rho W_\xi ],\ee from which the state can be
reobtained as
    $\rho= \int
    d^{2n} \xi\;
    \chi_\rho(\xi) W_\xi^\dagger/(2\pi)^n $.
Gaussian states are exactly those having  a Gaussian
characteristic function, and therefore a Gaussian Wigner function
in phase space:
\begin{equation}
    \chi_\rho (\xi) =
    e^{-  \xi^T \Gamma \xi/4 + D^T\xi
}.
\end{equation}
Here, the  $2n\times 2n$-matrix $\Gamma$ and the vector $D\in
\rr^{2n}$ are essentially the first and second moments: they are
related to the covariance matrix $\gamma$ and the displacements
$d$ as $\Gamma = \sigma^T \gamma \sigma$ and $D = \sigma d$. This
choice is then consistent with the definition of the {\it
covariance matrix} as having entries $
    \gamma_{j,k}=2 \text{Re}    \,
    \big\langle
    ( R_j - d_j )
    ( R_k - d_k )
    \big\rangle_\rho$,
$j,k=1,....,2n$, with $d_j={\rm tr}[R_j \rho ]$.
States always satisfy the Heisenberg uncertainty principle,
which can be expressed as $\gamma + i \sigma\geq 0$. This is a
simple semi-definite constraint onto any matrix of second moments,
 also obeyed by every non-Gaussian state.

\subsection{General Gaussian channels}

The simplest Gaussian channel is a lossless unitary evolution,
governed by a quadratic bosonic Hamiltonian:
\begin{equation}
    \rho\longmapsto U\rho U^\dagger,\quad U=e^{{\frac{i}2 \sum_{k,l}H_{kl} R_k
    R_l}},
\end{equation}
with $H$ being a real and symmetric $2n\times 2n$ matrix. Such
unitaries correspond to a representation of the real symplectic
group $Sp(2n,\rr)$, formed by those real matrices for which
$S\sigma S^T=\sigma$ \cite{book,Art,prama}. These are exactly the
linear transformations which preserve the commutation relations.
The relation between such a {\it canonical} transformation in
phase space and the corresponding unitary in Hilbert space is
given by $S=e^{H\sigma}$. Needless to say, Gaussian unitaries are
ubiquitous in physics, in particular in optics, and this is the
reason why Gaussian channels play such an important role. Notably,
the action of ideal beam splitters, phase shifters, and squeezers
correspond to symplectic transformations.\footnote{Any such $S$
can be decomposed into a {\it squeezing component}, and a {\it
passive operation} \cite{prama}. So one may write $S = V Z W$,
with $V, W\in K(n)= Sp(2n,\R)\cap SO(2n)$ are orthogonal
symplectic transformations, forming the subgroup of passive, i.e.,
number-preserving, operations. In turn, $Z=\,{\rm
diag}\,(z_1,1/z_1,\ldots,z_n,1/z_n)$ with $z_1,...,z_n\in \R
\backslash \{0\}$ are local single-mode squeezings.}

It is often instructive to consider  transformations on the level
of Weyl operators in the Heisenberg picture. For a symplectic
transformation we have $
    W_\xi \longmapsto W_{S^{-1} \xi}$.
The  action of a {\it general Gaussian channel} $\rho\longmapsto
T(\rho)$ can be phrased as
\begin{equation}
    W_\xi \longmapsto
    W_{X\xi} \;e^{-\frac12 \xi^T Y \xi},
\end{equation}
where $X,Y$ are real $2n\times2n$-matrices
\cite{Channels,Lindblad,book}. Additional linear terms in the
quadratic form are omitted since they merely result in
displacements in phase space, which are not interesting for later
purpose.  Not any transformation of the above form is possible:
complete positivity of the channel dictates that \footnote{The
case of a single mode is particularly transparent. Then, mixedness
can be expressed entirely in terms of determinants, and hence, the
above requirement can be cast into the form $Y\geq 0$, and
    $\det[ Y] \geq ( \det[ X] -1)^2$.}
\begin{equation}
    Y + i \sigma - i X^T \sigma X\geq 0.
\end{equation}
Depending on the context it may be more appropriate or transparent
to formulate a Gaussian channel in the Schr{\"o}dinger picture
$\rho\longmapsto T_{X,Y}(\rho)$ or to define it as a
transformation of  covariance matrices
\begin{equation}\label{TheOldMap}
    \gamma\longmapsto
    X^T\gamma X + Y.
\end{equation}

This is the most general form of a Gaussian channel. Roughly
speaking $X$ serves the purpose of amplification or attenuation
and rotation in phase space, whereas the $Y$ contribution is a
noise term which may consist of quantum (required to make the map
physical) as well as classical noise. Interestingly, $X$ may be
any real matrix, and hence, any map $\gamma\longmapsto X^T\gamma
X$ can be approximately realized, as long as 'sufficient noise' is
added. In this language, it also becomes immediately apparent how
much noise will be introduced by any physical device approximating
amplification or time reversal, meaning phase conjugation in an
optical context. For second moments far away from minimal
uncertainty, this additional noise may hardly have an impact (so
classical fields can be phase conjugated after all), whereas close
to minimal uncertainty this is not so any longer.

\subsection{Important examples of Gaussian channels}

The practically most important Gaussian channel is probably an
idealized action of a fiber. Moreover, as mentioned earlier, any
situation where a quadratic coupling to a Gaussian environment
provides a good description can be cast into the form of a
Gaussian channel. We will in the following consider a number of
important special cases of Gaussian channels for single modes:
\begin{enumerate}
\item The {\it classical noise channel} merely adds
classical Gaussian noise to a quantum state, i.e., $X=\id$, $Y\geq
0$ \cite{HolWe,HarringtonPreskill,Giovanetti,Ca}. In Schr\"odinger
picture this channel can be represented by a random displacement
according to a classical Gaussian probability distribution:
\begin{equation}
    T(\rho) = \frac1{4\pi\sqrt{\det Y}}
    \int d^2 \xi\;
     W_\xi \rho W_\xi^\dagger\;e^{-\frac14\xi^T Y^{-1}\xi}\;.
\end{equation}
\item  In the {\it thermal noise channel}
\cite{HolWe,Giovanetti} a mode passively interacts with another
mode in a thermal state, $
    \rho\longmapsto T(\rho)=\text{tr}_E[ U_\eta (\rho\otimes \rho_E) U_\eta^\dagger]
$. The result is as if the mode had been coupled in with a beam
splitter of some transmittivity $\eta$.\footnote{In the Heisenberg
picture this means that the annihilation operator transforms as
$a\mapsto\sqrt{\eta}\;a+\sqrt{1-\eta}\;b$, where $b$ is the
annihilation operator of the ancillary mode.} For the second
moments, we have that
\begin{equation}
    \gamma\longmapsto
    [S_\eta (\gamma \oplus c\id_2) S_\eta^T]_E,
\end{equation}
where $c\id_2$, $c\geq 1$, is the covariance matrix of a thermal
{\it Gibbs state}
\begin{equation}
    \rho_E = \frac{2}{c+1}\sum_{n=0}^\infty
    \left(
    \frac{c-1}{c+1}
    \right)^n |n\rangle\langle n|
\end{equation}
with mean photon number $(c-1)/2$. $[.]_E$ denotes the leading
$2\times 2$ submatrix. The passive symplectic transformation
$S_\eta$ is given by
\begin{equation}
    S_\eta=\left[
    \begin{array}{cc}
    \sqrt{\eta}\; \id_2 & \sqrt{1-\eta}\;\id_2\\
    - \sqrt{1-\eta}\;\id_2 & \sqrt{\eta}\; \id_2
    \end{array}
    \right],\,\,\eta\in[0,1].
\end{equation}
So we obtain
\begin{equation}\label{izf}
    \gamma\longmapsto
    \eta \gamma + (1-\eta) c\id_2.
\end{equation}
\item
The {\it lossy channel} is obtained by setting
$c=1$ in Eq(\ref{izf}). It reflects photon loss with probability
$1-\eta$. This channel is the prototype for optical communication
through a lossy fiber, since thermal photons (leading to a
contribution $c>1$) are negligible at room temperature. When using
an optical fiber of length $l$ and {\it absorbtion length} $l_A$
we may set $\eta=e^{-l/l_A}$. The lossy channel with $X=
\sqrt{\eta} \id_2$, $Y= (1-\eta)\id_2$ is also called {\it
attenuation channel} \cite{HolWe}.

\item The {\it
amplification channel} \cite{HolWe} is of the form
\begin{equation}
    X= \sqrt{\eta} \id_2 ,\,\,\,Y=(\eta -1)\id_2
    ,\quad\eta\in(1,\infty)\;.
\end{equation}
Here, the term $Y$ is a consequence of the noise that is added due
to Heisenberg uncertainty. Note that a classical noise channel can
be recovered as a concatenation of a lossy channel, followed by an
amplification.
\end{enumerate}

All these examples correspond to a single mode characterized by a
fixed frequency $\omega$. This is often referred to as the
narrowband case as opposed to {\it broadband channels}
\cite{Broadband,BroadbandPRL,Caves}, which consist out of many
uncoupled single-mode channels, each of which corresponds to a
certain frequency $\omega_i$, $i=1,2,...$ . Best studied is the
simple homogeneous case of a lossy broadband channel (equally
spaced frequencies $\omega_i$, with equal transmittivity $\eta$ in
all the modes).

It shall finally be mentioned that the very extensive literature
on harmonic open quantum systems is essentially concerned with
Gaussian channels of a specific kind, yet one where the
environment consists of infinitely many modes, where the linear
coupling is characterized by some spectral density.

\section{Entropies and quantum mutual information}\label{purity}

\subsection{Output entropies}

Channels describing the physical transmission of quantum states
typically introduce noise to the states as a consequence of a
decoherence process. Pure inputs are generally transformed into
mixed outputs, so into states $\rho$ having a positive von-Neumann
entropy
\begin{equation}
S(\rho)=-\text{tr}[\rho\log\rho].
\end{equation}
The entropy of the output will clearly depend on the input and the
channel itself, and the minimal such entropy can be taken as a
characteristic feature of the quantum channel. Introducing more
generally the $\alpha$ {\it Renyi entropies} for $\alpha\geq 0 $
as
\begin{equation}
    S_\alpha(\rho)
    = \frac{1}{1-\alpha}
    \log\text{tr}[ \rho^\alpha]
\end{equation}
this {\it minimal output entropy} \cite{amosov} is then defined
as\footnote{We use the notation $(S\circ
T)(\rho)=S\big(T(\rho)\big)$.}
\begin{equation}
    \nu_\alpha(T)
    =
    \inf_{\rho}
    (S_\alpha\circ T)(\rho ).
\end{equation}
The Renyi entropies \cite{Wehrl} are derived from the
$\alpha$-norms of the state, $ \|\rho \|_\alpha=  {\text{tr}}
[\rho ^\alpha]^{1/\alpha}$. In case of the limit
$\lim_{\alpha\searrow 1}$ one retains the von-Neumann entropy,
i.e., $ \lim_{\alpha\searrow 1}S_{\alpha}(\rho)= S(\rho)$; for
$\alpha=2$, this is the {\it purity} in the closer sense. Roughly
speaking, the smaller the minimal output entropy, the less
decohering is the channel (see, e.g., Ref.\ \cite{Il}).
The actual significance of this
quantity yet originates from its intimate relationship concerning
questions of capacities. This will be elaborated on in the
subsequent section.

\subsection{Mutual information and coherent information}

In Shannon's seminal channel coding theorem the capacity of a
classical channel is expressed in terms of the classical mutual
information \cite{Shannon}. In fact, as we will see below, the
quantum analogue of this quantity plays a similar role in quantum
information theory. For any quantum channel $T$ and any quantum
state $\rho$ acting on a Hilbert space $\cal H$, one defines the
{\it quantum mutual information} $I(\rho,T)$ as
\begin{equation}
    I(\rho,T) = S(\rho) + (S\circ T)(\rho) - S(\rho,T),
\end{equation}
where $S(\rho,T) = (\id \otimes T)(|\psi\rangle\langle\psi|)$ and
$|\psi\rangle\in{\cal H}_D\otimes{\cal H}$ is any purification of
the state $\rho=\text{tr}_D[|\psi\rangle\langle\psi|]$ \cite{HolWe,OldCerf}.
It is not
difficult to see that $I(\rho,T)$ does not depend on the chosen
purification. The quantum mutual information has many desirable
properties: it is positive, concave with respect to $\rho$, and
additive with respect to quantum channels of the form $T^{\otimes
n} $. The latter property  comes in very handy when relating this
quantity to the entanglement-assisted classical capacity.
An important part of the quantum mutual information is the {\it
coherent information} given by
\begin{equation}
    J(\rho,T) =   (S\circ T)(\rho) - S(\rho,T).
\end{equation}
 $J(\rho,T)$ can be
positive as well as negative, it is convex with respect to $T$ but
its convexity properties with respect to $\rho$ are unclear.

\subsection{Entropies of Gaussian states and extremal properties}

When maximizing the rate at which information can be sent through
a Gaussian channel, Gaussian states play an important role. In
fact, in many cases it turns out that encoding the information
into Gaussian states leads to the highest transmission rates. This
is mainly due to the fact that for a given covariance matrix
 many entropic quantities take on their extremal
values for Gaussian states. These entropic quantities, and in fact
any unitarily invariant functional, can for Gaussian states
immediately be read off the symplectic spectrum of the covariance
matrix: any covariance matrix $\gamma$ of $n$ modes can be brought
to the {\it Williamson normal form} \cite{williamson36},
$\gamma\longmapsto S\gamma S^T = \text{diag}(
c_1,c_1,c_2,c_2,...,c_n,c_n )$ with an appropriate $S\in
Sp(2n,\rr)$, and $\{c_i:i=1,...,n\}$ being the positive part of
the spectrum of $ i\sigma \gamma$. This is nothing but the
familiar normal mode decomposition with the $c_i$ corresponding to
the normal mode frequencies. Then, the problem of evaluating any
of the above quantities is reduced to a single-mode problem. For
example, the von-Neumann entropy is given by \cite{capa}
\begin{equation}
    S(\rho)=
    \sum_{i=1}^n
    g\Big(
    \frac{c_i-1}2
    \Big),
\end{equation}
where $g(N) =  (N+1)\log(N+1)-N\log N$ is the entropy of a thermal
Gaussian state with average photon number $N$. Similar expressions
can be found for the other entropic quantities.

Consider now any state $\tilde\rho$ which has the same first and
second moments as its Gaussian counterpart $\rho$. Then \be
S(\rho)-S(\tilde{\rho}) =
S(\tilde{\rho},\rho)+\tr\big[(\tilde{\rho}-\rho)\log\rho\big],\ee
where the first term is the nonnegative relative entropy, and the
second term vanishes since the expectation value of the operator
$\ln\rho$ depends only on the first and second moments. Hence, the
Gaussian state has the largest entropy among all states with a
given covariance matrix\footnote{The fact, that Gaussian states
maximize the entropy has far reaching consequences: (i) it is an
essential ingredient in showing that Gaussian input states achieve
the classical capacity for a lossy channel, (ii) it immediately
implies that bipartite states that contain the largest amount of
entanglement under an energy constraint are Gaussian, and (iii) it
implies that the entropy of a  Gaussian state is concave as a
function of the covariance matrix.} \cite{capa}. A more
sophisticated argument, using ideas of convex optimization and the
theorem of Kuhn and Tucker, shows that the same holds true for the
quantum mutual information \cite{HolWe}: For any Gaussian channel
$T$ and fixed first and second moments of $\rho$, the respective
Gaussian state maximizes $I(\rho,T)$. Whether a similar statement
also holds for the coherent information is not known.

Another very useful quantity that takes its
extremal values on Gaussian states, we would like to mention at this point,
is the {\it quantum conditional entropy} \cite{Negative}, defined as
\begin{equation}
        S(\rho: A|B)= S(\rho) - S(\rho_A)
\end{equation}
in a bi-partite system with parts $A$ and $B$. Here $\rho_A$ is
the reduction with respect to
system $A$. It can be shown in a very similar fashion as before
that this quantity is maximized on Gaussian states for fixed second moments,
although we now encounter a difference between two von-Neumann
entropies. Let $\tilde\rho$ again be a state with
the same first and second moments as its
Gaussian counterpart $\rho$, then
\begin{eqnarray}
        S(\rho: A|B) - S(\tilde\rho:A|B) &=&  S(\rho) - S(\rho_A)
        - S(\tilde\rho) + S(\tilde\rho_A)\nonumber\\
        &=& S(\tilde\rho|| \rho) - S(\tilde\rho_A|| \rho_A)\nonumber\\
        &+& \text{tr}[(\tilde\rho-\rho) \log\rho]
        -\text{tr}[(\tilde\rho_A-\rho_A) \log\rho_A]\geq 0.
\end{eqnarray}
In the last inequality, it is used that the relative entropy
can only decrease under joint application of completely positive maps.
This extremal property is helpful when assessing for example
achievable rates in state merging \cite{Negative}, for which the
quantum conditional entropy is an upper bound. More importantly
in the Gaussian setting, the negative of the conditional entropy
is a lower bound \cite{Hashing}
to the {\it distillable entanglement} \cite{Mother}, which
can be used to detect distillable entanglement in quantum
states by measuring second moments only. Whenever
one performs a measurement of second moments (estimation of the
variances of the quadratures) of an unknown state
$\tilde \rho$
and finds that its Gaussian counterpart satisfies
\begin{equation}
-S(\rho:A|B)= S(\rho_A) - S(\rho)>0,
\end{equation}
then one can argue that this is
in turn a lower bound for the distillable entanglement
$E_D(\tilde \rho)$ of $\tilde\rho$.
In this way, one can infer about the distillable entanglement
of an unknown quantum state
\begin{equation}
    E_D(\tilde \rho)\geq S(\rho_A) - S(\rho),
\end{equation}
without having to assume that the quantum state is Gaussian. This
is relevant as any knowledge whether a state is Gaussian is
typically not accessible without complete state tomography.
Moreover, this bound is robust against small perturbations, which
is also practically important since even complete state tomography
will determine the state only up to some error.

\subsection{Constrained quantities}

There are essentially two subtleties \cite{Subtle,Subtle2} that
arise in the infinite-dimensional context as we encounter it here
for Gaussian quantum channels: on the one hand, there is the
necessity of natural input constraints, such as one of finite mean
energy. Otherwise, the capacities diverge. On the other hand,
there is the possibility of continuous state
ensembles\footnote{This is understood as taking into account
probability measures on the set of quantum states.
For an approach in the language of probability and operator
theory, see Ref.\ \cite{Subtle}.}. The need for a constraint is
already obvious when considering the von-Neumann entropy: On a
state space over an infinite dimensional Hilbert space, the
von-Neumann entropy is not (trace-norm) continuous, but only lower
semi-continuous\footnote{This means that if, for a state $\rho$,
$\{\rho_n\}$ is a sequence of states for which $\rho_n\rightarrow
\rho$ in trace-norm as $n\rightarrow\infty$, then $S(\rho)\leq
\liminf_{n\rightarrow\infty} S(\rho_n)$.}, and almost everywhere
infinite.

This problem can be tamed by introducing an appropriate
constraint. For our purposes, we may take the Hamiltonian $H=
\sum_{i=1}^n (x_i^2 + p_i^2)/2$. Then, instead of taking all
states into account, one may consider the subset
\begin{equation}\label{TheConstraint}
    {\cal K} = \{\rho: \text{tr}[\rho H]<h\}.
\end{equation}
introducing for some $h>0$ a {\it constraint on the mean
energy}\footnote{More general constraints than this one can be
considered, leading to {\it compact subsets of state space} on
which one retains continuity properties in particular for the
von-Neumann entropy and the classical information capacity
\cite{Wehrl,Subtle,Subtle2,Infinity}. Essentially, any unbounded
positive operator $H$ with a spectrum without limiting points
would also be appropriate, such that $\text{tr}\exp[-\beta
H]<\infty$ for all $\beta>0$.} or mean photon number
$N=h-1/2$. Similarly, for tensor products we consider
${\cal K}^{\otimes n} = \{\rho: \text{tr}[\rho H^{\otimes n}]
< n h\}$. On this very natural subset ${\cal K}$ the
von-Neumann entropy and the classical information capacity retain
their continuity. In fact, many entanglement measures also retain
the continuity properties familiar in the finite-dimensional
context, such that, e.g., the entropy of a subsystem for pure
states can indeed be interpreted as the distillable entanglement
\cite{Infinity}.

\section{Capacities}

In classical information theory a single number describes how much
information  can reliably be sent through a channel: its {\it
capacity}. In quantum information theory the situation is more
complicated and each channel is characterized by a number of
different capacities \cite{Shor}. More precisely, which capacity
is the relevant one depends on whether we want to transmit
classical or quantum information, and on the resources and
protocols we allow for. An important resource that we must
consider is entanglement shared between sender and receiver. The
presence or absence of this resource together with the question
about sending classical or quantum information leads to four basic
capacities, which we will discuss in the following.

\subsection{Classical information capacity}

The {\it classical information capacity} is the asymptotically
achievable number of classical bits that can be reliably
transmitted from a sender to a receiver per use of the channel.
Here, it is assumed that the parties may coherently encode and
decode the information in the sense that they may use entangled
states as codewords at the input and joint measurements over
arbitrary channel uses at the output. This answers essentially the
question of how useful a quantum channel is for the transmission
of classical information.

This capacity is derived from the single-shot expression
\cite{OldHolevo,OldWest}, appropriately constrained as above,
\begin{equation}\label{SS}
        C_1(T,{\cal K})=
        \sup
        \left[
        S\bigl(\sum_i p_i T(\rho_i)\bigr)
        - \sum_i p_i (S\circ T) (\rho_i)
        \right],
\end{equation}
where the supremum is taken over all probability distributions and
sets of states satisfying $\rho=\sum_i p_i \rho_i$ under the
constraint $\rho\in {\cal K}$ \cite{Subtle,Subtle2} \footnote{The
above constraint also ensures that $(S\circ T)(\rho)<\infty$.
The convex hull function of $S\circ T$, given by
 $\rho\longmapsto \hat S(\rho,T)=
\inf \sum_i p_i (S\circ T)(\rho_i)$ in Eq.\ (\ref{SS}), with the
infimum being taken over all ensembles with $\sum_i p_i \rho_i
=\rho$, is still convex in the unconstrained case, but no longer
continuous, however, lower semi-continuous in the above sense.}.
By the Holevo-Schumacher-Westmoreland (HSW) theorem
\cite{OldHolevo,OldWest}, this single-shot expression gives the
capacity if the encoding is restricted to product states. Hence,
the full classical information capacity can formally be expressed
as the regularization
of $C_1$,
\begin{equation}\label{DefCc}
        C(T, {\cal K}) = \lim_{n\rightarrow \infty}
        \frac{1}{n} C_1(T^{\otimes n},{\cal K}^{\otimes n}).
\end{equation}
Clearly, $C(T, {\cal K})\geq C_1(T,{\cal K})$ since the latter
does not allow for inputs which are entangled over several
instances of the channel. Yet, it is in general not known whether
this possibility comes along with any advantage at all, so whether
entangled inputs facilitate a better information transfer. This
will be remarked on later.

Note that in this infinite-dimensional setting, the constraint is
required to obtain a meaningful expression for the capacity: for
all non-trivial Gaussian channels the optimization over all input
ensembles in Eq.\ (\ref{SS}) would lead to an infinite capacity.
This can simply be achieved by encoding the information into phase
space translates of any signal state. Then no matter how much
noise is induced by the channel, we can always choose the spacing
between the different signal states sufficiently large such that
they can be distinguished nearly perfectly at the output.

Let us now follow the lines of Ref.\ \cite{ClassicalCapacity} and
sketch the derivation of the classical capacity for lossy
channels. First of all, a lower bound on $C(T,{\cal K})$ can be
obtained by choosing an explicit input ensemble for Eq.\
(\ref{SS}). Random coding over coherent states according to a
classical Gaussian probability distribution leads to an average
input state of the form \be \rho\propto\int d^2\xi\; W_\xi
|0\rangle\langle 0|W_\xi^\dagger\; e^{-\frac14
\xi^TV^{-1}\xi}\;,\ee with covariance matrix $\gamma=\id+V$.
Hence, if we choose $V=2N\id$, the average number of photons in
the input state will be $\tr[\rho a^\dagger a]=N$. The constraint
set ${\cal K}$ hence corresponds to the choice of $h=N+1/2$. After
passing a lossy channel with transmittivity $\eta$ this changes to
$\tr[T(\rho) a^\dagger a]=\eta N$, and since $T(\rho)$ is a
thermal state, its entropy is given by $(S\circ T)(\rho) =g(\eta
N) $. The action of a lossy channel on a coherent input state is
to shift the state by a factor $\eta$ towards the origin in phase
space. In other words, the channel maps coherent states onto
coherent states and since the latter have zero entropy, we have
\cite{HolWe}
\begin{equation}
    C_1(T,{\cal K})\geq (S\circ T)(\rho)=g(\eta N).
\end{equation}

Assume now that $\tilde\rho$ is the average input state optimizing
$C_1(T^{\otimes n},{\cal K}^{\otimes n})$ under a given constraint
for the mean energy as described above. Then
\begin{equation}
    C_1(T^{\otimes n},{\cal K}^{\otimes n})
    \leq (S\circ T^{\otimes n}) (\tilde{\rho})
    \leq
    \sum_{i=1}^n (S\circ T)(\tilde{\rho}_i)\;,
\end{equation}
where $\tilde{\rho}_i$ is the reduction of $\tilde\rho$ to the
$i$-th mode and the second inequality is due to the subadditivity
of the von-Neumann entropy. Since for a fixed average photon
number $\tr[\tilde{\rho}_i a^\dagger a]=N_i$ the entropy is
maximized by a Gaussian state, we have in addition that $(S\circ
T) (\tilde{\rho}_i)\leq g(\eta N_i)$.

Together with the lower bound this implies that the classical
capacity of a lossy channel is indeed given by $C(T,{\cal
K})=g(\eta N)$ \cite{ClassicalCapacity}, if the average number of
input photons per channel use is restricted to be not larger than
$N$, corresponding to the constrained associated with ${\cal K}$.
Hence random coding over coherent states turns out to be optimal
and neither non-classical signal states nor entanglement is
required in the encoding step.\footnote{Of course, there might
also be optimal encodings which do exploit a number state alphabet
or entanglement between successive channel uses.}

An immediate consequence of this result is that the classical
capacity of the homogeneous broadband channel $T$ is given by \be
C(T,{\cal K})=t\frac{\sqrt{\eta}}{\ln 2}\sqrt{\frac{\pi
P}{3}}+{\cal O}(1/t)\;,\ee where $P$ is the average input power
and $t$ is the transmission time related to the frequency spacing
$\delta\omega=2\pi/t$. For the lossless case $\eta=1$ this
capacity was derived in Ref. \cite{Caves,YO}.

\subsection{Quantum capacities}

The {\it quantum capacity} is the rate at which qubits can be
reliably transmitted through the channel from a sender to a
receiver. This transmission is done again employing appropriate
encodings and decodings before envoking instances of the quantum
channel \cite{HolWe}. This capacity can be made precise using the
{\it norm of complete boundedness} \footnote{This is defined as
$||T||_{\text{cb}}=\sup_n ||\text{Id}_n\otimes T||$, where
$||T||=\sup_X ||T(X)||_1/||X||_1$.}. The question is how well the
identity channel can be approximated in this norm. More
specifically \cite{dennis1}, the quantum capacity $Q$ is the
supremum of $c\geq 0$ such that for all $\varepsilon,\delta>0$
there exist $n,m\in \nn$, decodings $T_{\text{D}}$ and encodings
$T_{\text{E}}$ with
\begin{equation}
    \left|
    \frac{n}{m} - c
    \right|<\delta,\,\,\,\,\,
    \left\|\text{Id}^{\otimes n}_2 - T_{\text{D}}  T^{\otimes m}
    T_{\text{E}} \right\|_{\text{cb}}< \varepsilon.
\end{equation}
One may also consider a weaker instance, allowing for
$\varepsilon$-errors, and then look at a
$Q_{\varepsilon}$-capacity \cite{HolWe}. It is known that the
quantum capacity does not increase if we allow for additional
classical forward communication \cite{littlehelp}.

In Ref.\ \cite{Qtheorem} it was proven that the quantum capacity $Q(T)$
can be expressed in terms of the coherent information as \be
Q(T)=\lim_{n\rightarrow\infty} \frac1n \sup_\rho
J\big(\rho,T^{\otimes n}\big)\;.\ee Unfortunately, the asymptotic
regularization is required in general, since the supremum over the
coherent information is known to be not additive\footnote{Note
also that while the subtleties in the infinite-dimensional context
have been fleshed out and precisely clarified for the classical
information capacity \cite{Subtle,Subtle2}, the
entanglement-assisted capacity \cite{Constrained}, and measures of
entanglement \cite{Infinity,Subtle2}, questions of continuity
related to the quantum capacity are to our knowledge still
awaiting a rigorous formulation.}. However, the single-shot
quantity $\sup_\rho J(\rho,T)$ already gives a useful lower bound
on $Q(T)$. For the classical noise Gaussian channel and
Gaussian $\rho$ this was first shown to be attainable in Ref.\
\cite{HarringtonPreskill}, based on earlier work \cite{Gottesman},
using methods of quantum stabilizer codes that embed a
finite-dimensional protected code space in an infinite-dimensional
one.
For more general thermal noise channels, this is given by
\cite{HolWe} \bea \label{JD} J(\rho,T)&=&
g(N')-g\left(\frac{D+N'-N-1}2\right)-g\left(\frac{D-N'+N-1}2\right),\;\
\ \quad \\
D&=&\sqrt{(N+N'+1)^{2 }-4\eta N(N+1)}\;, \eea where $N'=\eta
N+(1-\eta)(c-1)/2$ is the average photon number at the channel
output. In fact, the same bound holds for the amplification
channel, for which $\eta>1$ and $N'=\eta N+(\eta-1)(c+1)/2$. For
broadband channels, lower bounds of this kind on the quantum
capacity were discussed in Ref.\ \cite{Broadband}.

A computable upper bound on the quantum capacity of any channel is
given by $Q(T)\leq \log||T\theta||_{\text{cb}}$ \cite{HolWe}. For
finite-dimensional systems $\theta$ is the matrix transposition,
which corresponds to the momentum-reversal operation in the
continuous variables case. This bound is zero for {\it
entanglement breaking} channels\footnote{A channel is called
entanglement breaking if it corresponds to a measure and
repreparation scheme.} and additive for tensor products of
channels. For attenuation and amplification channels with
classical noise, i.e., channels acting as $\gamma\mapsto
\eta\gamma+|1-\eta|c$, this leads to \cite{HolWe}
\begin{equation}
    Q(T)\leq
    \log(1+\eta)-
    \log|1-\eta|-\log c\;.
\end{equation}
Note that this bound is finite for all $\eta\neq 1$. This is
remarkable since it does not depend on the input energy. That is,
unlike the classical capacity, the unconstrained quantum capacity
does typically not diverge. Moreover, it is even zero in the case
$\eta\leq 1/2$, since then the no-cloning theorem forbids an
asymptotic error-free transmission of quantum information.

\subsection{Entanglement-assisted  capacities}

Needless to say, in a quantum information context, it is
meaningful to see what rates can be achieved for the transfer of
classical information when entanglement is present. This is the
kind of information transfer considered in the {\it
entanglement-assisted classical capacity} $C_E$
\cite{Assisted,Constrained}. It is defined as the rate at which
bits that can be transmitted in a reliable manner in the presence
of an unlimited amount of prior entanglement shared between the
sender and the receiver. In just the same manner, the {\it
entanglement-assisted quantum capacity} $Q_E$ may be defined
\cite{Constrained,Broadband,BroadbandPRL}. Similarly, this
quantifies the rate at which qubits can asymptotically be reliably
transmitted per channel use, again in the presence of unlimited
entanglement. Exploiting teleportation and dense coding is not
difficult to see that $2 Q_E= C_E$. Now, the entanglement-assisted
capacity $C_E$ is intimately related to the
 quantum mutual information, as just the supremum of this
quantity with respect to all states $\rho\in {\cal K}$ as in Eq.\
(\ref{TheConstraint})
\begin{equation}\label{CE}
    C_E(T,{\cal K}) = \sup_\rho I(\rho,T).
\end{equation}
Again, with this constraint \cite{Constrained}, the quantity
regains the appropriate continuity properties\footnote{In a more
general formulation -- i.e., for non-Gaussian constrained
channels, or for Gaussian channels with different constraints  --
one has to require that $\sup_{\rho\in{\cal K}} (S\circ T)
(\rho)<\infty$ \cite{Constrained}.}. Note that in this case, no
asymptotic version has to be considered, and due to the additivity
of the quantum mutual information the single-shot expression
already provides the capacity.

In a sense Eq.\ (\ref{CE}) is the direct analogue of Shannon's
classical coding theorem. The latter states that the classical
capacity of a classical channel is given by the maximum mutual
information. The main difference is, however, that in the
classical case shared randomness does not increase the capacity,
whereas for quantum channels shared entanglement typically
increases the capacity,
\begin{equation}
    C(T,{\cal K})\leq C_E(T,{\cal K}).
\end{equation}
Again, similar to the classical case $C_E$ is conjectured to characterize
equivalence classes of channels within which all channels can
efficiently simulate one another \cite{Assisted}.

For Gaussian channels the extremal property of Gaussian states
with respect to the quantum mutual information allows us to
calculate $C_E(T,{\cal K})$ by only maximizing over constrained
Gaussian states $\rho$. For attenuation channels with classical
noise, i.e., $\gamma\mapsto \eta\gamma+(1-\eta)c$ with
$0\leq\eta\leq 1$, it was shown in Ref.\ \cite{HolWe} that \be
C_E(T,{\cal K})=g(N)+J(\rho,T)\;,\ee with the coherent information
$J(\rho,T)$ taken from Eq.\ (\ref{JD}). For the homogeneous
broadband lossy channel, extensively discussed in Ref.\
\cite{Broadband,BroadbandPRL}, it holds again that $C_E(T,{\cal
K})\propto t\sqrt{P}$.

\section{Additivity issues}

In the previous sections, we have encountered additivity problems
of several quantities related to quantum channels. Such questions
are at the core of quantum information theory: essentially, the
questions is whether for product channels one can potentially
gain from utilizing entangled inputs. This applies in particular
to the additivity of the single-shot expression $C_1$ and the
minimal output entropy\footnote{In the context of entanglement
measures, additivity refers to the property that for a number of
uncorrelated bi-partite systems, the degrees of entanglement
simply add up to the total entanglement.} \cite{amosov}. A number
of partial results on additivity problems have been found. Yet, a
conclusive answer to the most central additivity questions is
still lacking. In particular, it is one of the indeed intriguing
open questions of quantum information science whether the
single-shot expression $C_1$ in Eq.\ (\ref{SS}) is already
identical to the classical information capacity as it is true for
the case of the lossy channel \cite{ClassicalCapacity}.

\subsection{Equivalence of additivity problems}

Interestingly, a number of additivity questions are related in the
sense that they are either all true or all false. This connection
is particularly well-established in the finite-dimensional context
\cite{Equivalence,matsu,Koen}: then, the equivalence of the (i)
additivity of the minimum output $1$-entropy, the von-Neumann
entropy, (ii) the additivity of the single-shot expression $C_1$,
(iii) the additivity of the entanglement of formation, and (iv)
the strong superadditivity of the entanglement of formation have
been shown to be equivalent \cite{Equivalence,matsu,Koen}. This
equivalence, besides being an interesting result in its own right,
provides convenient starting points for general studies on
additivity, as in particular the minimal output entropies appear
much more accessible than the classical information capacity.

In the infinite-dimensional context, the argument concerning
equivalence is somewhat burdened with technicalities. We will here
state the main part of an equivalence theorem of additivity
questions concerning any pair $T_1$, $T_2$ of Gaussian channels
\cite{Subtle2}. The following
properties (1.) and (2.) are equivalent and imply (3.):\\

\begin{enumerate}
    \item[(1.)]
For any state $\rho$ on the product Hilbert space and for all appropriately constraint sets
${\cal K}_1$ and ${\cal K}_2$
we have that\footnote{This has to hold for all compact subsets
${\cal K}_1$ and ${\cal K}_2$ of state space for which $(S\circ
T_i)(\rho)<\infty$ for all states $\rho\in {\cal K}_i$, $i=1,2$,
and such that $C_1 (T_1,{\cal K}_1),C_1 (T_2,{\cal K}_2)<\infty$.
Note that these assumptions are in particular satisfied if ${\cal
K}_1$ and ${\cal K}_2$ are defined by an energy constraint.}
\begin{equation}
    C_1 (T_1\otimes T_2, {\cal K}_1\otimes {\cal K}_2)
    = C_1 (T_1,{\cal K}_1) + C_1 (T_2,{\cal K}_2).
\end{equation}

    \item[(2.)] For any state $\rho$ with $(S\circ
T_1)(\text{tr}_2[\rho])<\infty$ and $(S\circ
T_2)(\text{tr}_1[\rho])<\infty$
\begin{equation}
    \hat S(\rho,T_1\otimes T_2) \geq
    \hat S(\text{tr}_2[\rho], T_1)
    +
    \hat S(\text{tr}_1[\rho], T_2),
\end{equation}
where for a channel $T$ and $\sum_i p_i\rho_i = \rho$
\begin{equation}
    \hat S(\rho,T) =
    \inf \sum_i p_i (S\circ T)(\rho_i).
\end{equation}
    \item[(3.)] For the minimal output entropies
\begin{equation}
    \bar \nu_1(T_1\otimes T_2) =
    \nu_1(T_1) + \nu_1 (T_2)
\end{equation}
where the bar indicates that in order to evaluate the minimal
output entropy of $T_1\otimes T_2$, the infimum is taken only
over all pure states $\rho$ such that $S(\text{tr}_2[\rho])=
S(\text{tr}_1[\rho])<\infty$ and $(S\circ (T_1\otimes T_2))(\rho)
<\infty$. \\
\end{enumerate}

 In particular, this means that
once a general answer to (1.) or (2.) was known for Gaussian
channels, a general single-shot expression for the classical
information capacity of such channels would be available, solving
a long-standing open question. Moreover, it was proven that the
above implications hold true if one of the additivity conjectures
is proven for the general finite dimensional case \cite{Subtle2}.

\subsection{Integer output entropies}

For specific channels, the unconstrained minimal output
$\alpha$-entropies for tensor products can be identified for
integer $\alpha$. These {\it integer instances of output purities}
are not immediately related to the question of the classical
information capacity, for which the limit $\alpha\searrow 1$ is
needed. However, they provide a strong indication of additivity
 also in the general case. Notably, for the single-mode classical
and thermal noise channels $T$,
\begin{equation}
        \nu_\alpha(T^{\otimes n}) = n \nu_\alpha(T)
\end{equation}
 has been established for integer $\alpha$ \cite{Giovanetti}. The
concept of entrywise positive maps also provides a general
framework for assessing integer minimal output entropies for
Gaussian channels \cite{Entrywise}, generalizing previous results.
It is worth mentioning that in the above cases the minimal output
entropy $\nu_\alpha(T)$, $2\leq\alpha\in\nn$ is attained
for Gaussian input states \cite{Giovanetti}.

\subsection{Output entropies for Gaussian inputs}

In all known cases Gaussian input states achieve the minimal
output entropy or attain the capacity of Gaussian channels. Hence,
one may be tempted to believe that this could be true in general
and thus consider only Gaussian input states from the very
beginning. In this restricted settings, quite far-reaching
statements concerning additivity can yet be made. For example, if
one requires that the encoding is done entirely in Gaussian terms,
the additivity for minimal output entropies can be proven in quite
some generality \cite{OurGaussian}. The {\it Gaussian minimal
output entropy} is defined as
\begin{equation}
    \nu_{\alpha,G}(T) = \inf_\rho (S_\alpha\circ T)(\rho),
 \end{equation}
where the infimum is taken over all Gaussian states. Then one
finds that the minimal output $\alpha$-entropy for single-mode
Gaussian channels $T_1,...,T_n$, as in Eq.\ (\ref{TheOldMap})
characterized by $X_1,...,X_n$ and $Y_1,...,Y_n$, $Y_i\geq 0$, and
$\text{det}[X_i]=\text{det}[X_j]$ for all $i,j$ is additive for
all $\alpha\in (1,\infty)$. This includes the important case of
identical Gaussian channels $T$,
\begin{equation}
    \nu_{\alpha,G}(T^{\otimes n}) = n \nu_{\alpha,G}(T)
\end{equation}
for all $n$ and all $\alpha\in(1,\infty)$. Moreover, for
$\alpha=2$ this kind of additivity was proven for arbitrary
multi-mode Gaussian channels for which $\det[X_i]\neq 0$ \cite{OurGaussian}.

\subsection{Equivalence of Gaussian additivity problems}

The aim of this section is to make a first step towards proving
that the equivalence of additivity problems
\cite{Equivalence,matsu,Koen} holds within the Gaussian world,
where all resources and quantities are appropriately replaced by
their Gaussian counterparts. Since all states in this section are Gaussian
and thus essentially characterized by their covariance matrices
(except from the first moments, which are not relevant, e.g.,
for their entanglement content and their entropy), we will for
simplicity of notation
use the covariance matrix $\gamma$ as the argument of
functions which are supposed to act on density operators. That
is, we will write $T(\gamma)$ and $S(\gamma)$ meaning
$T(\rho_\gamma)$ and $S(\rho_\gamma)$, with $\rho_\gamma$ being
the centered Gaussian state with covariance matrix $\gamma$.
Let us first define the quantities under consideration:

\begin{enumerate}

\item[I.] {\it Gaussian entanglement of formation:} The
Gaussian version \cite{GEOF} of the {\it entanglement of
formation} (EoF) \cite{Mother}
restricts to decompositions into Gaussian states
-- the probability distribution in the decomposition is however
not restricted to be Gaussian (although there exists always an
optimal decomposition with this property \cite{GEOF}). As proven
in Ref.\ \cite{GEOF}, we have
\begin{eqnarray}
    E_G(\gamma)& =&\inf_{\det[\Gamma]=1}\Big\{E(\Gamma)\big|\gamma\geq\Gamma
    \geq
    E_G(\gamma)\Bigr\} \nonumber \\
    &=&\inf_{\det[\Gamma]=1}\Big\{E(\Gamma)\big|\gamma\geq\Gamma\geq
    i\sigma\Big\},\label{eq:GEoF}
\end{eqnarray}
 where $E(\Gamma)$ is the entropy of
entanglement of the pure Gaussian state with covariance matrix
$\Gamma$, i.e., the von-Neumann entropy of its reduced state.
The corresponding decomposition contains only
phase-space displaced versions of this state. Obviously this is an
upper bound for the (unconstrained) {\it entanglement of formation}
\cite{Mother},
i.e., $E_G(\gamma)\geq
E_F(\gamma)$ where equality holds at least for the case of
symmetric two-mode states (for which the reduced states are
   unitarily equivalent) \cite{EOF}. For these states $E_G$ was
proven to be additive \cite{GEOF}
\be
    E_G\big(\oplus_{i=1}^n
    \gamma_{i}\big)=\sum_{i=1}^n
    E_G(\gamma_i)=\sum_{i=1}^n
    E_F(\gamma_i)\;,
\ee
and convex on the level of covariance
matrices, i.e.,
\be\label{convex}
    E_G\big(\lambda\gamma_1+(1-\lambda)\gamma_2\big)\leq \lambda
    E_G(\gamma_1)+(1-\lambda)E_G(\gamma_2)\;
\ee
for all $\lambda\in[0,1]$.

    \item[II.] {\it Gaussian capacity:}
We introduce the Gaussian counterpart $C_{1,G}(T,{\cal K})$
     of the single-shot expression
$C_{1}(T,{\cal K})$ in Eq.\ (\ref{SS})
by restricting the input ensemble to be a set of phase-space translates
     of a Gaussian state, distributed according to a Gaussian distribution.
       Evidently, $C_{1,G}(T,{\cal K})\leq
C_{1}(T,{\cal K})$ and the question of additivity is, whether for
all Gaussian channels equality holds in
    \be \label{GC}\sum_{i=1}^n
    C_{1,G}(T_i,{\cal K}_i)\leq
    C_{1,G}\Big(\otimes_{i=1}^n
    T_i,\;\otimes_{i=1}^n {\cal K}_i\Big)\;.
\ee
As mentioned above, for the lossy channel we have indeed that \cite{ClassicalCapacity}
\begin{equation}
    C_{1,G}(T,{\cal K})=C_{1}(T,{\cal K})=C(T,{\cal K}).
\end{equation}
\end{enumerate}

{\it Gaussian MSW correspondence}: Following  Matsumoto, Shimono,
and Winter (MSW) in Ref.\
\cite{matsu}, one can easily establish a relation
between $E_G$, $C_{1,G}$, and $\nu_{1,G}$. Let $T:\gamma\mapsto
    X\gamma X^T+Y$ be a Gaussian channel acting on systems of
    $n$-modes. Then there exists a dilation, i.e., a pure state of
    $m\leq 2n$ modes with covariance matrix $\gamma_0$ and a
      symplectic transformation $S$, such that
\be\label{eq:dilation}
    T(\gamma)=[\gamma']_A\ ,\quad
    \gamma' =S(\gamma\oplus\gamma_0)S^T,\ee
where $A$ and $B$ refer to an $n$-mode and $m$-mode subsystem,
and as before $[\gamma']_A$
denotes the covariance matrix corresponding to subsystem $A$.
This is nothing but the corresponding principal submatrix.
Replacing $\cal K$ by the singleton set of a fixed average
Gaussian input state with covariance matrix  $\gamma$ leads to a new
quantity
$C_{1,G}(T,\gamma)$, which is defined as
\be\label{eq:MSW1}
    C_{1,G}(T,\gamma)= (S\circ T) (\gamma)-E_G\big(\gamma'\big)\;.
\ee
This relates the Gaussian EoF to the capacity $C_{1,G}$. In fact,
if ${\cal G}(T,{\cal K})$ is the set of all covariance matrices
$\gamma'$ in Eq.\ (\ref{eq:dilation}) for which $\gamma\in{\cal
K}$, then \be C_{1,G}(T,{\cal K})=\sup_{\Gamma\in {\cal G}(T,{\cal
K})} S\big([\Gamma]_A\big)-E_G\big(\Gamma\big),\ee which is the
Gaussian analogue of the MSW correspondence \cite{matsu}. Moreover,
the simplicity of the Gaussian EoF in Eq.\ (\ref{eq:GEoF}) leads to
a relation between $E_G$ and the minimum output entropy: if
$\gamma,\gamma'$, and $T$ are again related via
Eq.\ (\ref{eq:dilation}), then
\bea\label{eq:EGHminG}
    E_G\big(\gamma'\big) &=&
    \inf_{i\sigma\leq\tilde{\gamma}\leq\gamma}
    (S\circ T)(\tilde{\gamma}),\\
    \nu_{1,G}(T) &=&
    \inf_{\gamma} E_G(\gamma')\;.
\eea

{\it Implications for Gaussian additivity problems}:
Using the above
Gaussian analogue of the MSW correspondence and following the
argumentation in Ref.\ \cite{Equivalence,matsu},
one can easily prove
that Gaussian versions of all the above additivity statements
would be implied by the super-additivity of the Gaussian
entanglement of formation. Let $\gamma$ be the covariance matrix
of a bi-partite
Gaussian state consisting of $n$ bi-partite
sub-systems\footnote{Each sub-system may in turn consist out of an
arbitrary (but finite) number of modes, jointly forming sub-systems $A$ and $B$.}
with respective reduced covariance matrices $[\gamma]_i$.
Then $E_G$ is said to be {\it super-additive} if
\be\label{suppe}
    E_G(\gamma)\geq \sum_{i=1}^n
    E_G([\gamma]_i)\;.
\ee
Note that $\gamma$ is not assumed to be of direct sum structure.
If Eq.\ (\ref{suppe}) holds for
all covariance matrices $\gamma$,
then
\begin{enumerate}
    \item[(I.)] $E_G$ is additive, i.e.,
    $E_G(\gamma_1\oplus\gamma_2)=E_G(\gamma_1)+E_G(\gamma_2)$,
    \item[(II.)]  the constrained Gaussian classical capacity is additive,
    i.e., equality holds in Eq.\ (\ref{GC}),
    \item[(III.)]  the minimal output entropy restricted to Gaussian inputs
    is additive, i.e., $\nu_{1,G}(T_1\otimes
    T_2)=\nu_{1,G}(T_1)+\nu_{1,G}(T_1)$,
    \item[(IV.)]
    $E_G$ is convex on the level of covariance matrices.
\end{enumerate} Here, (I.) is evident, and (II.), (III.) are proven
in close analogy to Refs.\ \cite{Equivalence,matsu}. Statement
(IV.) is shown as follows: consider two bi-partite covariance
matrices $\gamma_1$ and $\gamma_2$ of equal size. There is a local
symplectic transformation $S$ (consisting out of 50:50 beam
splitters), which
acts as \be\label{bls} S(\gamma_1\oplus\gamma_2)S^T=\frac12\left(%
\begin{array}{cc}
      \gamma_1+\gamma_2 & \gamma_1-\gamma_2 \\
    \gamma_1-\gamma_2 & \gamma_1+\gamma_2 \\
    \end{array}%
    \right)\;=:\; \Gamma\;.\ee By the implied additivity of $E_G$
   and its unitary
    invariance\footnote{$E_G\big(S(\gamma_1\oplus\gamma_2)S^T\big)=E_G(\gamma_1\oplus\gamma_2)$
    since $S$ is a local unitary.}, we have that
    $E_G(\gamma_1)+E_G(\gamma_2)=E_G(\Gamma)$. Moreover,
super-additivity implies that $E_G(\Gamma)\geq
2E_G\big((\gamma_1+\gamma_2)/2\big)$, resulting in convexity for
the case $\lambda=1/2$. By interpolation and continuity this can
then be extended to the entire interval $\lambda\in[0,1]$.

Remarkably, this implication has a simple converse: if $E_G$ is
additive and convex on the level of covariance matrices, then it is
super-additive. To see this, we introduce a local symplectic
transformation $\theta=\id\oplus(-\id)$ with block structure as in
Eq.\ (\ref{bls}). Then, for every
\begin{equation}
    \Gamma= {{\left(
    \begin{array}{cc}
       \gamma_1 & C \\
       C^T & \gamma_2 \\
    \end{array}
    \right)}}
\end{equation}
we have
\bea E_G(\Gamma) &=&\Big[
E_G(\Gamma)+\ E_G(\theta\Gamma\theta)\Big]/2\nonumber\\
&\geq& E_G\big((\Gamma+\theta\Gamma\theta)/2\big) =
E_G(\gamma_1\oplus\gamma_2)\nonumber\\
&=& E_G(\gamma_1)+E_G(\gamma_2)\;,\eea where the inequality is due
to the assumed convexity and the last equation reflects additivity
of $E_G$.
Note that by the above result, if $E_G$ is not convex on covariance matrices,
then either $E_F\neq E_G$ or $E_F$ is not additive.

\section{Outlook}

This article was concerned with the theory of Gaussian quantum
communication channels. Such channels arise in several practical
contexts, most importantly as models for lossy fibers. Emphasis
was put on questions related to capacities, which give the best
possible bounds on the rates that can be achieved when using
channels for the communication of quantum or classical
information.

Though many basic questions have been solved over the last few
years, many interesting questions in the theory of bosonic
Gaussian channels are essentially open. This applies in particular
to additivity issues: general formulae for the classical
information capacity are simply not available before a resolution
of these issues. For specific channels, a number of methods can
yet be applied to find additivity of output purities. It may be
interesting to see how far the idea of relating minimal
$1$-entropies to $2$-entropies as in Ref.\ \cite{Additive} could
be extended in the infinite-dimensional context.

Then, there is the old conjecture that to take Gaussian ensembles
does not constitute a restriction of generality anyway when
transmitting information through a Gaussian quantum channel. In
the light of this conjecture, it would be interesting whether a
complete theory of quantum communication can be formulated,
restricting both to Gaussian ensembles and Gaussian channels.

Finally, all what has been stated on capacities in this article
refers to the case of memoryless channels. For Gaussian channels
with memory, the situation can be quite different. For example,
notably, the classical information capacity can be enhanced using
entangled instead of product inputs \cite{Cerf,Memory2}. It would
in this context also be interesting to see the program of Ref.\
\cite{NewMemory} implemented in the practically important case of
Gaussian quantum channels.

\section*{Acknowledgments}
\addcontentsline{toc}{section}{Acknowledgements}

We would like to thank D.\ Kretschmann for helpful comments
on the manuscript. 
This work has been supported by the EPSRC (GR/S82176/0, QIP-IRC),
the European Union (QUPRODIS, IST-2001-38877), the DFG
(Schwerpunktprogramm QIV), and the European Research Councils
(EURYI).


\bigskip


\begin{thebibliography}{99}

\bibitem{Channels}
        B.\ Demoen, P.\ Vanheuswijn, and A.\ Verbeure,
        Lett.\ Math.\ Phys.\ {\bf 2}, 161 (1977).

\bibitem{capa}
        A.S.\ Holevo, M.\ Sohma, and O.\ Hirota, Phys.\ Rev.\ A {\bf 59}, 1820
        (1999).

\bibitem{HolWe}
        A.S.\  Holevo and R.F.\ Werner, Phys.\ Rev.\ A {\bf 63}, 032312 (2001).

\bibitem{OurOldChannels}
        J.\ Eisert and M.B.\ Plenio, Phys.\ Rev.\ Lett.\ {\bf 89}, 097901
        (2002).

\bibitem{Lindblad}
        G.\ Lindblad,
        J.\ Phys.\ A {\bf 33}, 5059 (2000).

\bibitem{Holevobook}
    A.S.\ Holevo, {\it Probabilistic Aspects of
    Quantum Theory} (North-Holland, Amsterdam, 1982), Chapter 5.

\bibitem{book}
    J.I.\ Cirac, J.\ Eisert, G.\ Giedke,  M.B.\ Plenio, M.\
        Lewenstein,
        M.M.\ Wolf, and R.F.\ Werner,
        textbook in preparation (2005).

\bibitem{Art}
        J.\ Eisert and M.B.\ Plenio,
        Int.\ J.\ Quant.\ Inf.\ {\bf 1},  479 (2003).

\bibitem{prama}
    Arvind, B.\ Dutta, N.\ Mukunda, and R.\ Simon,
    Pramana {\bf 45}, 471 (1995); quant-ph/9509002.

\bibitem{HarringtonPreskill}
        J.\ Harrington and J.\ Preskill,
        Phys.\ Rev.\ A   {\bf 64}, 062301 (2001).

\bibitem{Giovanetti}
        V.\ Giovannetti, S.\ Lloyd, L.\ Maccone, J.H.\ Shapiro, and B.J.\ Yen,
        Phys.\ Rev.\ A {\bf 70}, 022328 (2004).

\bibitem{Ca}
        C.M.\ Caves and K.\ Wodkiewicz, quant-ph/0409063.

\bibitem{Broadband}
     V.\ Giovannetti, S.\ Lloyd, L.\ Maccone, and P.W.\ Shor,
     Phys.\ Rev.\ A {\bf 68}, 062323 (2003).

\bibitem{BroadbandPRL}
    V.\ Giovannetti, S.\ Lloyd, L.\ Maccone, and P.W.\ Shor,
        Phys.\ Rev.\ Lett. {\bf 91}, 047901 (2003).

\bibitem{Caves}
        C.M.\ Caves and P.D.\ Drummond,
        Rev.\ Mod.\ Phys.\ {\bf 66}, 481  (1994).

\bibitem{amosov}
        G.G.\ Amosov, A.S.\ Holevo, and R.F.\
        Werner, Problems
        in Information Transmission {\bf 36}, 25 (2000).

\bibitem{Wehrl}
    A.\ Wehrl, Rev.\ Mod.\ Phys.\ {\bf 50}, 221 (1978).


\bibitem{Il}
        A.\ Serafini, F.\ Illuminati, M.G.A.\ Paris, and S.\ De Siena,
        Phys.\ Rev.\ A {\bf 69}, 022318 (2004).

\bibitem{OldCerf}
    C.\ Adami and N.J.\ Cerf, Phys.\ Rev.\ A {\bf 57}, 4153 (1998).

\bibitem{Shannon}
    C.E.\ Shannon, The Bell System Tech.\ J.\ {\bf 27},
    379 (1948); ibid.\ {\bf 27}, 623 (1948).

\bibitem{williamson36}
        J.\ Williamson, Am.\ J.\ Math.\ {\bf 58}, 141 (1936); see also
        V.I.\ Arnold, {\em Mathematical Methods of Classical Mechanics},
        (Springer-Verlag, New York,
        1978).

\bibitem{Negative}
     M.\ Horodecki, J.\ Oppenheim, and A.\ Winter,
    quant-ph/0505062 (2005).

\bibitem{Hashing}
     I.\ Devetak and A.\ Winter,
    Proc.\ R.\ Soc.\ Lond. A {\bf  461}, 207 (2005).

\bibitem{Mother}
     C.H.\ Bennett, D.P.\ DiVincenzo, J.A.\ Smolin, and W.K.\ Wootters,
    Phys.\ Rev.\ A {\bf 54}, 3824 (1996).

\bibitem{Subtle}
        A.S.\ Holevo and M.E.\ Shirokov, quant-ph/0408176 (2004).

\bibitem{Subtle2}
    M.E.\ Shirokov, quant-ph/0411091 (2004).

\bibitem{Infinity}
    J.\ Eisert, C.\ Simon, and M.B.\ Plenio,
    J.\ Phys.\ A {\bf 35}, 3911 (2002).

\bibitem{Shor}
        P.W.\ Shor, Math.\ Prog.\ {\bf 97}, 311 (2003).

\bibitem{OldHolevo}
    A.S.\ Holevo, IEEE Trans.\ Inf.\ Theory {\bf 44}, 269 (1998).

\bibitem{OldWest}
    B.\ Schumacher and M.D.\ Westmoreland, Phys.\ Rev.\ A {\bf 56}, 131
    (1997).

\bibitem{ClassicalCapacity}
     V.\ Giovannetti, S.\ Guha, S.\ Lloyd, L.\ Maccone,
    J.H.\ Shapiro, and H.P.\ Yuen,
    Phys.\ Rev.\ Lett.\ {\bf 92}, 027902 (2004).

\bibitem{YO}
    H.P.\ Yuen and M.\ Ozawa, Phys.\ Rev.\ Lett.\ {\bf 70},
    363 (1992).

\bibitem{dennis1}
    D.\ Kretschmann and R.F.\ Werner, New J.\ Phys.\ {\bf 6}, 26
    (2004).

\bibitem{littlehelp}
    C.H.\ Bennett, D.P.\ DiVincenzo, J.A.\
    Smolin, and W.K.\ Wootters, Phys.\ Rev.\ A {\bf 54}, 3824 (1996);
    H.\ Barnum, E.\ Knill, and M.A.\ Nielsen, IEEE
    Trans.\ Inf.\ Th.\ {\bf
    46}, 1317 (2000).

\bibitem{Qtheorem}
    P.W.\ Shor, {\it The quantum channel capacity
    and coherent information}, lecture notes, MSRI Workshop on Quantum
    Computation (2002); I. Devetak, IEEE Trans.\ Inf.\ Th.\ {\bf 51}, 44 (2005);
    S.\ Lloyd,
    Phys.\ Rev.\ A {\bf 55}, 1613 (1997).

\bibitem{Constrained}
        A.S.\ Holevo, quant-ph/0211170 (2002).

\bibitem{Gottesman}
        D.\ Gottesman, A.\ Kitaev, and J.\ Preskill,
        Phys.\ Rev.\ A {\bf 64}, 012310  (2001).

\bibitem{Assisted}
    C.H.\ Bennett, P.W.\ Shor, J.A.\ Smolin, A.V.\
    Thapliyal, IEEE Trans Inf.\ Th.\
    {\bf 48}, 2637 (2002).

\bibitem{Equivalence}
        P.W.\ Shor,
    Comm.\ Math.\ Phys.\ {\bf 246}, 453 (2004).

\bibitem{matsu}
        K.\ Matsumoto, T.\ Shimono, and A.\ Winter,
    Commun.\ Math.\ Phys.\ {\bf 246}, 427 (2004).

\bibitem{Koen}
        K.M.R.\ Audenaert and S.L.\ Braunstein,
        Commun.\ Math.\ Phys.\ {\bf 246},
        443 (2004).

\bibitem{OurGaussian}
        A.\ Serafini, J.\ Eisert, and M.M.\ Wolf,
        Phys.\ Rev.\ A {\bf 71}, 012320 (2005).

\bibitem{GEOF}
    M.M.\ Wolf, G.\ Giedke, O.\ Krueger, R.F.\ Werner, and J.I.\ Cirac,
     Phys.\ Rev.\ A {\bf 69}, 052320 (2004).

\bibitem{EOF}
    G.\ Giedke, M.M.\ Wolf, O.\ Krueger, R.F.\ Werner, and J.I.\ Cirac,
    Phys.\ Rev.\ Lett.\ {\bf 91}, 107901 (2003).

\bibitem{Entrywise}
        C.\ King, M.\ Nathanson, and M.B.\ Ruskai, quant-ph/0409181 (2004).

\bibitem{Additive}
        M.M.\ Wolf and J.\ Eisert,
        New J.\ Phys.\ {\bf 7}, 93 (2005).

\bibitem{Cerf}
        N.J.\ Cerf, J.\ Clavareau, C.\ Macchiavello, and J.\ Roland,
        quant-ph/0412089 (2004).

\bibitem{Memory2}
    G.\ Ruggeri, G.\ Soliani, V.\ Giovannetti, and S.\ Mancini,
    quant-ph/0502093 (2005).

\bibitem{NewMemory}
    D.\ Kretschmann and R.F.\ Werner, quant-ph/0502106 (2005).



\end{thebibliography}
\end{document}